\documentclass[twocolumn, secnumarabic, amssymb, nobibnotes, aps, prb, superscriptaddress]{revtex4-2}

\usepackage{amsmath}

\usepackage[final]{graphicx}

\begin{document}

\title{
 Static and Dynamic Electronic Properties and the Possible Magnetic Structure of the $4f^{3}$-$\Gamma_{6}$ System NdCo$_{2}$Zn$_{18}$Ga$_{2}$ Investigated Using $^{59}$Co Nuclear Quadrupole Resonance
}

\author{Tetsuro~Kubo}%
\altaffiliation{tetsurok111@gmail.com}
\affiliation{Department of Physics, Graduate School of Science, Kobe University, Kobe 657-8501, Japan}

\author{Atsushi~Sasaki}%
\affiliation{Department of Physics, Graduate School of Science, Kobe University, Kobe 657-8501, Japan}

\author{Keita~Murooka}%
\affiliation{Department of Physics, Graduate School of Science, Kobe University, Kobe 657-8501, Japan}

\author{Hisashi~Kotegawa}%
\affiliation{Department of Physics, Graduate School of Science, Kobe University, Kobe 657-8501, Japan}

\author{Rikako~Yamamoto}%
\altaffiliation{Present address: Max Planck Institute for Chemical Physics of Solids, 01187 Dresden, Germany}
\affiliation{Department of Quantum Matter, Graduate School of Advanced Science and Engineering, Hiroshima University, Higashi-Hiroshima 739-8530, Japan}

\author{Takahiro~Onimaru}%
\affiliation{Department of Quantum Matter, Graduate School of Advanced Science and Engineering, Hiroshima University, Higashi-Hiroshima 739-8530, Japan}

\author{Hideki~Tou}%
\altaffiliation{tou@crystal.kobe-u.ac.jp}
\affiliation{Department of Physics, Graduate School of Science, Kobe University, Kobe 657-8501, Japan}

\date{
\today
}%

\begin{abstract}
We report $^{59}$Co nuclear quadrupole resonance (NQR) measurements on the Nd-based compound NdCo$_{2}$Zn$_{18}$Ga$_{2}$, which undergoes an antiferromagnetic transition at $T_{\rm N} = 1.5$~K. Although the NQR spectra show no detectable change across $T_{\rm N}$, the nuclear spin-lattice relaxation rate, $1 / T_{1}$, exhibits a clear anomaly at $T_{\rm N}$. An analysis based on the alignment of the Nd moments demonstrates that the internal magnetic fields generated by these moments cancel each other at the Co sites. If the nearest-neighbor Nd moments align antiferromagnetically, this finding suggests that Ga substitution removes magnetic frustration, thereby increasing $T_{\rm N}$.
\end{abstract}

\maketitle

The rich physical phenomena arising from electron correlations have become a central theme in condensed matter physics. In particular, metallic compounds containing rare-earth or actinide elements with $f$ electrons exhibit heavy-fermion behavior, unconventional superconductivity, and multipolar ordering through interactions between $f$ and conduction electrons.

Among thses materials, $R T_{2} X_{20}$ compounds ($R$ = rare earth, $T$ = transition metals, $X$ = Zn, Al, Cd, or Mg) have attracted considerable interest because of their diverse physical properties. These properties are attributed to the high symmetry at the rare-earth ion sites and the effective enhancement of hybridization owing to the cage-like structure. In Pr$T_{2} X_{20}$ compounds, quadrupolar ordering, non-Fermi-liquid states, and superconductivity have been observed~\cite{Onimaru2010, Onimaru2011, Sakai2011, Sakai2012, Onimaru2012, Matsubayashi2012, Tsujimoto2014, Machida2015, Onimaru2016, Yoshida2017, Fu2020}, while YbCo$_{2}$Zn$_{20}$ hosts an extremely heavy-electron state originating from pseudo-degenerate crystal-field states~\cite{Torikachvili2007}.

Within this family, Nd$T_{2} X_{20}$ compounds with a $4f^{3}$ configuration exhibit magnetic ordering at low temperatures owing to Kramers degeneracy. Single-crystal studies have been conducted on Nd$T_{2}$Zn$_{20}$ ($T =$ Co, Rh, and Ir), where the competition between antiferromagnetic and ferromagnetic interactions between $4f$ moments has attracted interest from the perspective of magnetic frustration between the first- and second-nearest-neighbor Nd moments~\cite{Yamane2017, Yamamoto2019, Yamamoto2022}.

NdCo$_{2}$Zn$_{20}$ has a $\Gamma_{6}$ doublet crystalline-electric-field (CEF) ground state and undergoes a first-order antiferromagnetic transition at $T_{\rm N} = 0.53$~K~\cite{Yamamoto2019,Yamamoto2023}. For the $4f^{3}$ configuration of Nd$^{3+}$, Hotta proposed that a magnetic two-channel Kondo effect can emerge from a $\Gamma_{6}$ doublet based on the renormalization group analysis of a seven-orbital impurity Anderson model hybridized with conduction electrons of $\Gamma_{8}$ symmetry. This results in a residual entropy of $0.5 R \ln 2$~\cite{Hotta2017}. This residual entropy is a hallmark of the non-Fermi liquid behavior associated with the two-channel Kondo effect, wherein conduction electrons overscreen the localized moments. The relevance of either the two-channel Kondo effect or the magnetic frustration inherent in the diamond structure has been suggested for NdCo$_{2}$Zn$_{20}$ based on entropy release and the convex-upward temperature dependence of the electrical resistivity between $T_{\rm N}$ and 4~K~\cite{Yamamoto2019}.

To clarify the ground state properties and magnetic interactions in NdCo$_{2}$Zn$_{20}$, the effects of chemical substitution have been investigated. Sn substitution, which introduces a negative chemical pressure, reduces CEF splitting~\cite{Ishikawa2015}. Studies on Cd and Ga substitutions have also been conducted. In Cd-substituted systems, the disorder in the exchange interactions between $4f$ electrons broadens the magnetic transition~\cite{Yamamoto2021}.

Of particular interest is Ga substitution in NdCo$_{2}$Zn$_{20 - y}$Ga$_{y}$. From comparisons with other substitution systems~\cite{Ishikawa2015,Wakiya2018}, it is believed that Ga selectively occupies the Zn $16c$ site. For $y = 1$ and 2, Ga substitution increases the lattice constant and $T_{\rm N}$ to 0.78 and 1.5~K, respectively, while changing the transition from first- to second-order. These changes suggest that $4p$-electron doping strengthens the antiferromagnetic interactions~\cite{Yamamoto2021}. Notably, for $y = 2$, $T_{\rm N}$ triples, and the transition becomes second order and is well reproduced by mean-field calculations, implying that the anomalous magnetic state observed in the non-substituted system has been suppressed.

However, the magnetic structure of these Ga-substituted systems remains unclear. Therefore, we investigate the magnetic interactions and possible magnetic structures of the Ga-substituted NdCo$_{2}$Zn$_{18}$Ga$_{2}$ ($y = 2$) using $^{59}$Co nuclear quadrupole resonance (NQR). Nuclear magnetic resonance (NMR) and NQR techniques are powerful tools for probing local electronic and magnetic properties, allowing us to examine the nature of magnetic ordering and dynamics at Co sites.

Polycrystalline NdCo$_{2}$Zn$_{18}$Ga$_{2}$ was prepared using the vertical Bridgman method. The details of the sample preparation have been reported elsewhere~\cite{Yamamoto2021}. For the measurements, we used a coarse-grained sample obtained by crushing polycrystals.

$^{59}$Co-NQR measurements (nuclear spin $I = 7/2$, gyromagnetic ratio $^{59}\gamma / 2 \pi = 10.102130$~MHz/T, and nuclear quadrupole moment $^{59}Q = 0.42 \times 10^{-28}$~m$^{2}$) were performed using a conventional pulsed-NMR method with a phase-coherent spectrometer. An 8~T superconducting magnet with a $^{4}$He cryostat was used for $T > 1$~K, while a $^{3}$He-$^{4}$He dilution refrigerator was used for $T < 1$~K. NMR and NQR spectra were obtained by integrating the spin echoes at each frequency. The nuclear spin-lattice relaxation rate $1 / T_{1}$ was measured using an inversion recovery method.

\begin{figure}[h]
 \centering
 \includegraphics[width=1.00\linewidth]{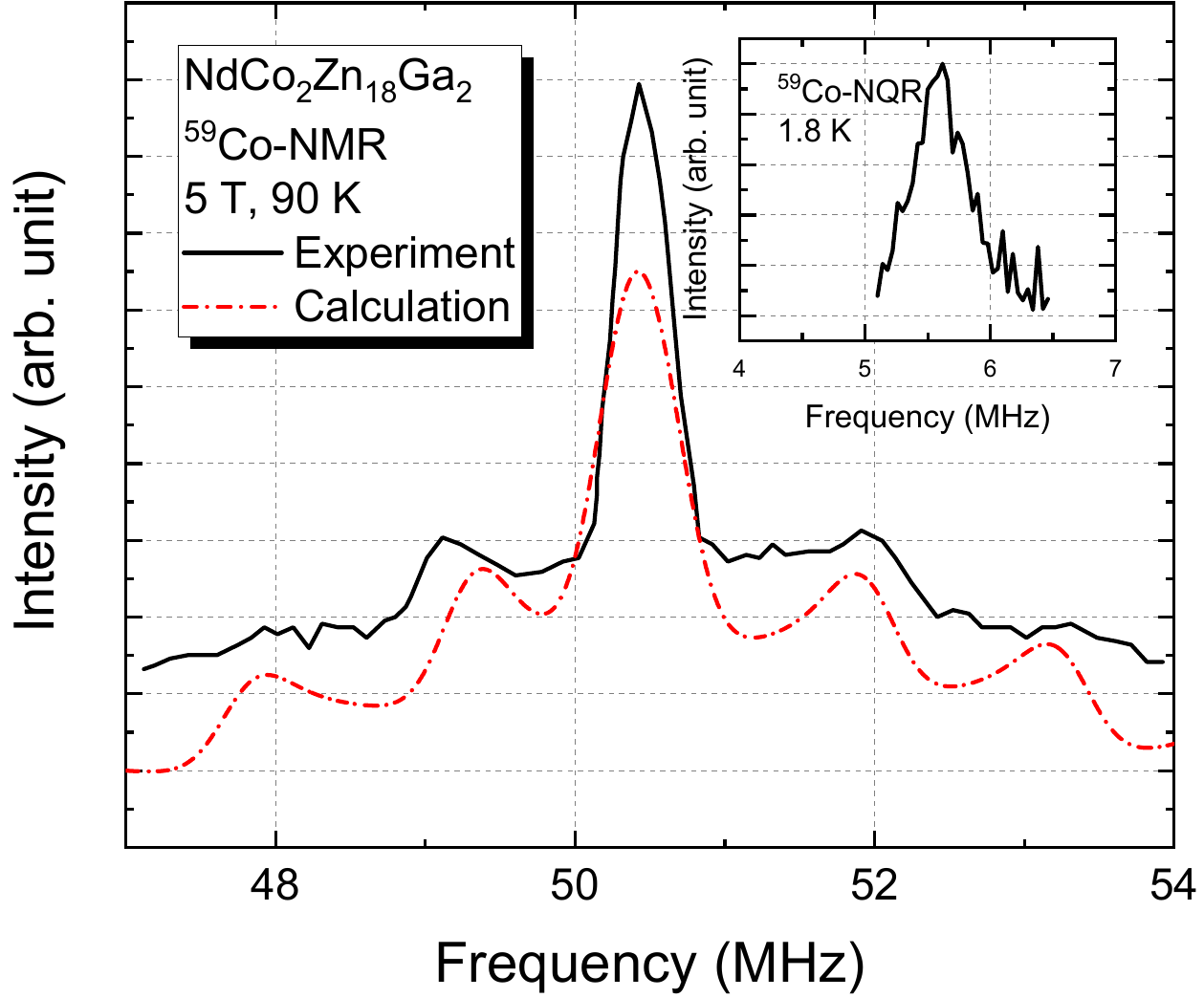}
 \caption{(Color Online) $^{59}$Co-NMR powder spectrum of NdCo$_{2}$Zn$_{18}$Ga$_{2}$ obtained at $B = 5$~T and $T = 90$~K. The dashed curve represents a powder-pattern calculation with $K = 0.9$~\%, $\nu_{\rm Q} = 2.8$~MHz, and $\eta = 0$. The inset displays the $^{59}$Co-NQR $2 \nu_{\rm Q}$ transition at 1.8~K.}
 \label{fig:NMRspectrum}
\end{figure}

Figure~\ref{fig:NMRspectrum} shows the $^{59}$Co-NMR spectrum obtained at $B = 5$~T and $T = 90$~K for NdCo$_{2}$Zn$_{18}$Ga$_{2}$. The spectrum exhibits the characteristic features of a powder pattern with quadrupolar splitting. The Hamiltonian for the $^{59}$Co nuclear spin can be expressed as
\begin{align}
 \mathcal{H} & = \mathcal{H}_{\rm Z} + \mathcal{H}_{\rm Q} \\
 \mathcal{H}_{\rm Z} & = - \gamma_{\rm n} \hbar (1 + K) {\bf I} \cdot {\bf H}_{\rm ext} \\
 \mathcal{H}_{\rm Q} & = \frac{1}{6} h \nu_{\rm Q} \left\{3 I_{Z}^{2} - I (I + 1) + \frac{1}{2} \eta \left(I_{+}^{2} + I_{-}^{2}\right)\right\} \\
 \nu_{\rm Q} & = \frac{3 e Q V_{ZZ}}{2 I (2 I - 1) h}, \quad \eta = \frac{V_{XX} - V_{YY}}{V_{ZZ}},
\end{align}
where $\gamma_{\rm n}$ is the nuclear gyromagnetic ratio, $K$ is the Knight shift, $\nu_{\rm Q}$ is the nuclear quadrupole resonance frequency, and $\eta$ is the asymmetry parameter of the electric field gradient (EFG). Here, $V_{\alpha \alpha}$ ($\alpha = X, Y, Z$) are the principal EFG values at the Co site. Figure~\ref{fig:NMRspectrum} shows the results of a powder pattern simulation performed at $K = 0.9$\%, $\nu_{\rm Q} = 2.8$~MHz, and $\eta = 0$. The simulation reproduces the shape of the experimental line. Based on the estimated $\nu_{\rm Q}$ value, we performed NQR measurements and observed three NQR lines. The inset of Fig.~\ref{fig:NMRspectrum} shows a representative $2 \nu_{\rm Q}$ line spectrum at 1.8~K. Notably, $\nu_{\rm Q}$ increased by approximately 40\% upon Ga substitution compared with nonmagnetic reference compounds such as YCo$_{2}$Zn$_{20}$~\cite{Kubo2025} and LuCo$_{2}$Zn$_{20}$~\cite{Mito2013}. Similar $\nu_{\rm Q}$ values have been reported for Sn-substituted (La,Pr)Co$_{2}$Zn$_{18}$Sn$_{2}$~\cite{Magishi2020}, suggesting that $p$-electron doping through substitution at the Zn cage modifies the EFG around the Co sites. The full width at half maximum (FWHM) of the $2 \nu_{\rm Q}$ line was approximately 0.5~MHz, whereas the FWHM of the isostructural compounds LuCo$_{2}$Zn$_{20}$ and YbCo$_{2}$Zn$_{20}$ were approximately 50~kHz~\cite{Mito2013}. The relatively broad linewidth in the present system may be attributed to the local environmental effects caused by Ga substitution.

\begin{figure}[h]
 \centering
 \includegraphics[width=1.00\linewidth]{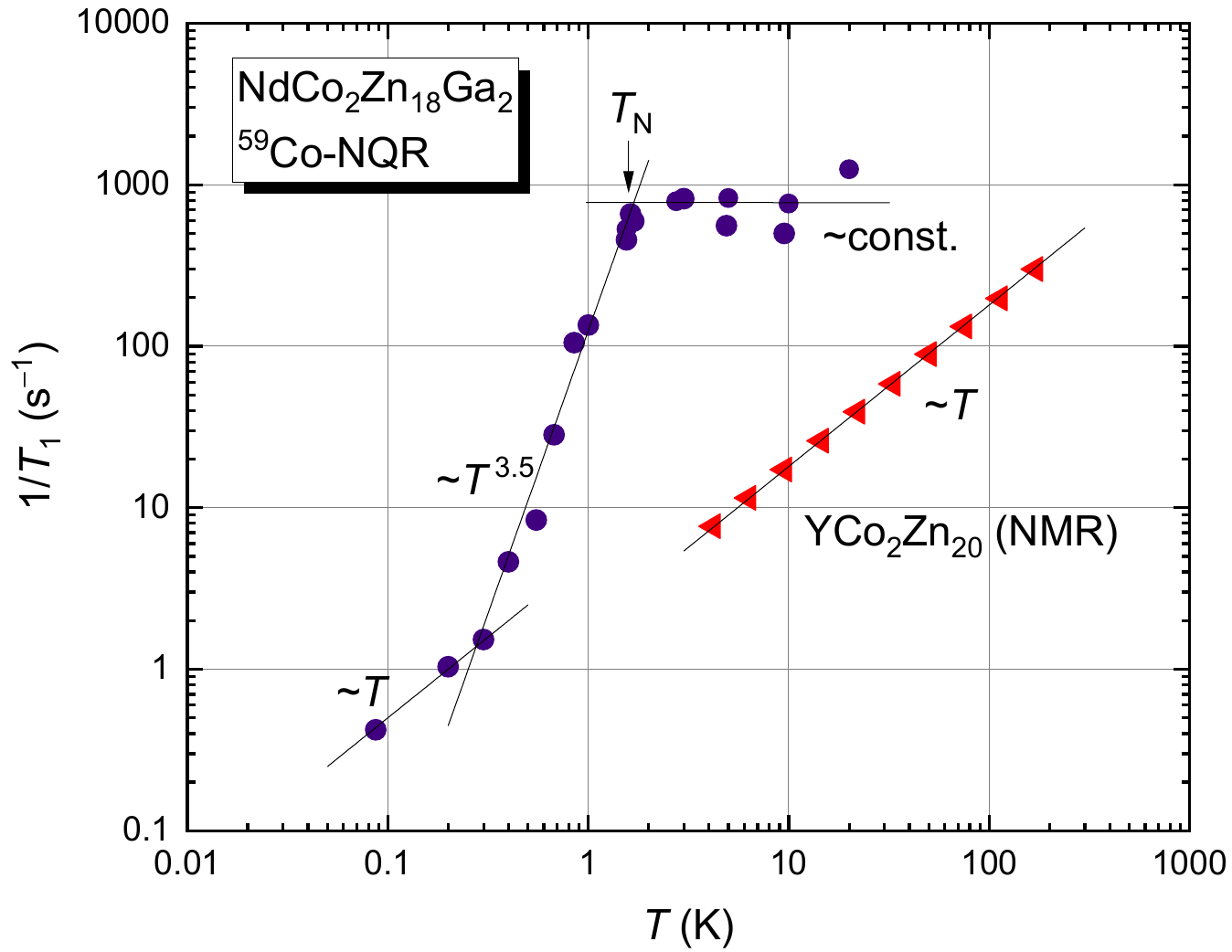}
 \caption{(Color Online) Temperature dependence of $1 / T_{1}$ for NdCo$_{2}$Zn$_{18}$Ga$_{2}$ (NQR) and YCo$_{2}$Zn$_{20}$ ($^{59}$Co-NMR at 7~T). The solid lines are visual guides.}
 \label{fig:T1T}
\end{figure}

Figure~\ref{fig:T1T} shows the temperature dependence of the NQR nuclear spin-lattice relaxation rate $1 / T_{1}$ for NdCo$_{2}$Zn$_{18}$Ga$_{2}$, along with that of the nonmagnetic reference compound YCo$_{2}$Zn$_{20}$ measured at 7~T. The $1 / T_{1} T$ value of YCo$_{2}$Zn$_{20}$ follows the Korringa relation, $1 / T_{1} \sim T$, between 4 and 200~K, indicating typical metallic behavior.

By contrast, $1 / T_{1}$ for NdCo$_{2}$Zn$_{18}$Ga$_{2}$ remained nearly constant between 1.5 and 20~K, suggesting that nuclear relaxation is dominated by fluctuations of localized Nd moments. Below 1.5~K, $1/T_{1}$ exhibited a sharp decrease, indicating a phase transition. This temperature coincides with bulk $T_{\rm N}$, providing microscopic evidence of the transition. For $T < T_{\rm N}$, $1 / T_{1}$ follows a $\sim T^{3.5}$ dependence. In antiferromagnets below $T_{\rm N}$, $1/T_{1}$ is known to follow distinct temperature dependences: $1/T_{1} \sim T^{3}$ for $T \gg T_{\rm AE}$ and $1/T_{1} \sim T^{2} \exp(- T_{\rm AE} / T)$ for $T \ll T_{\rm AE}$, where $T_{\rm AE}$ represents the spin-wave excitation gap~\cite{Jaccarino}. In the antiferromagnetic insulator UO$_{2}$, $1/T_{1} \sim T^{7}$ behavior arising from magnon-phonon excitations has been observed for $T < T_{\rm N}$~\cite{Ikushima2001}. As we experimentally observed power-law behavior, we suggest that a partial gap opens in the magnon excitation spectrum upon antiferromagnetic ordering. Alternatively, the decrease in $1 / T_{1}$ can be explained by a partial loss of the density of states owing to the opening of an energy gap below $T_{\rm N}$; for instance, the folding of the Fermi surface caused by the formation of the magnetic unit cell. At the lowest temperatures, where $1 / T_{1} \sim T$, nuclear relaxation is governed primarily by conduction electrons.

\begin{figure}[h]
 \centering
 \includegraphics[width=1.00\linewidth]{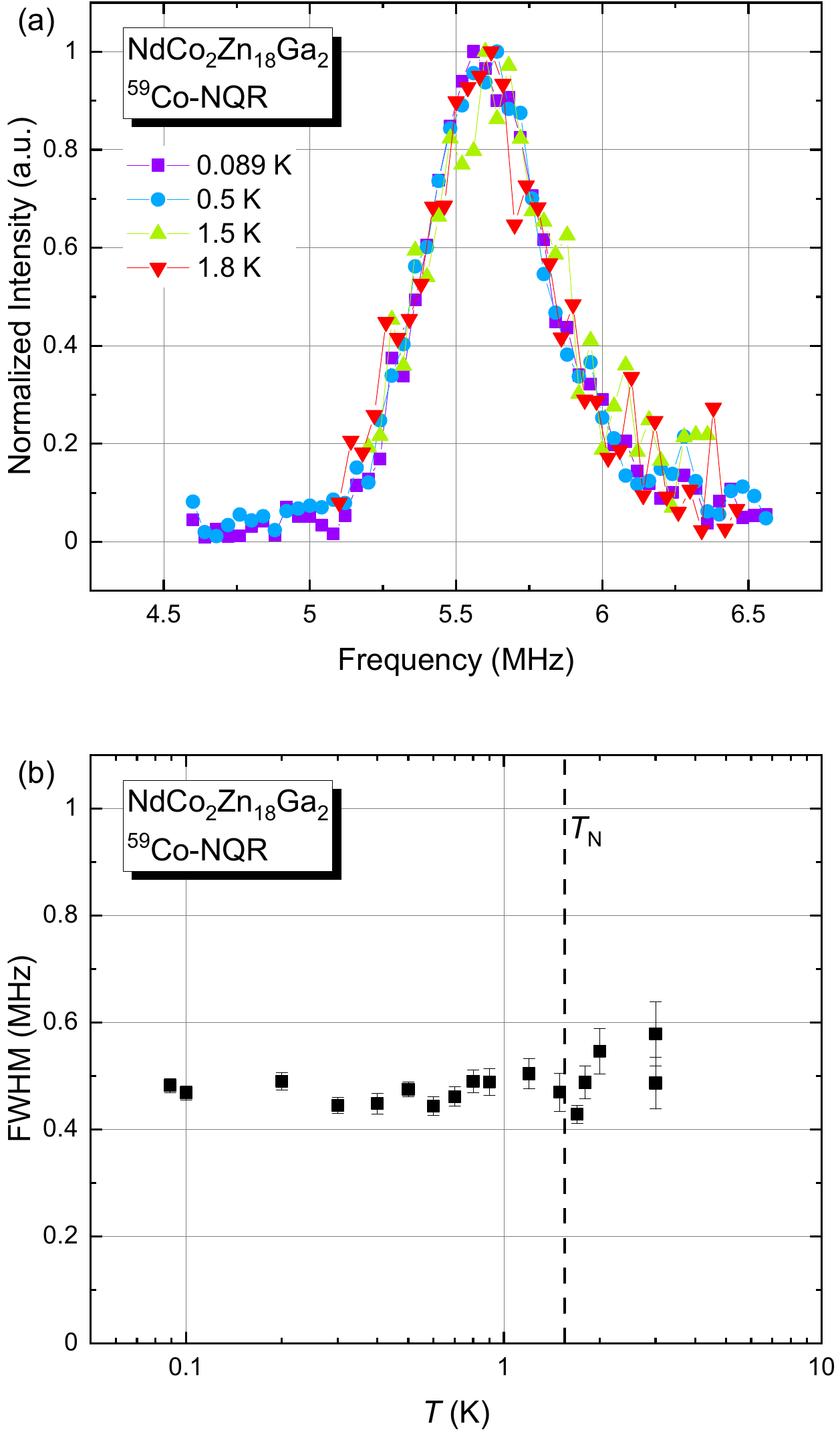}
 \caption{(Color Online) Temperature dependence of (a) the NQR line shape and (b) the full width at half maximum (FWHM) of the $2 \nu_{\rm Q}$ transition in NdCo$_{2}$Zn$_{18}$Ga$_{2}$.}
 \label{fig:NQRspectrum}
\end{figure}

The temperature dependence of the NQR spectra from 0.089 to 1.8~K is shown in Fig.~\ref{fig:NQRspectrum}(a) and the corresponding FWHM values are plotted in Fig.~\ref{fig:NQRspectrum}(b). No spectral splitting or noticeable broadening was observed across $T_{\rm N}$.

The magnetic ordering in NdCo$_{2}$Zn$_{18}$Ga$_{2}$ is next discussed based on the NQR measurements. The anomaly in $1 / T_{1}$ at $T_{\rm N}$ is consistent with previous bulk measurements. However, the absence of peak shifts or line broadening in the NQR spectra suggests that the static hyperfine fields created by the Nd magnetic moments cancel each other out at the Co sites. This microscopic evidence confirms that the transition at $T_{\rm N}$ is of antiferromagnetic origin.

\begin{figure}[h]
 \centering
 \includegraphics[width=1.00\linewidth]{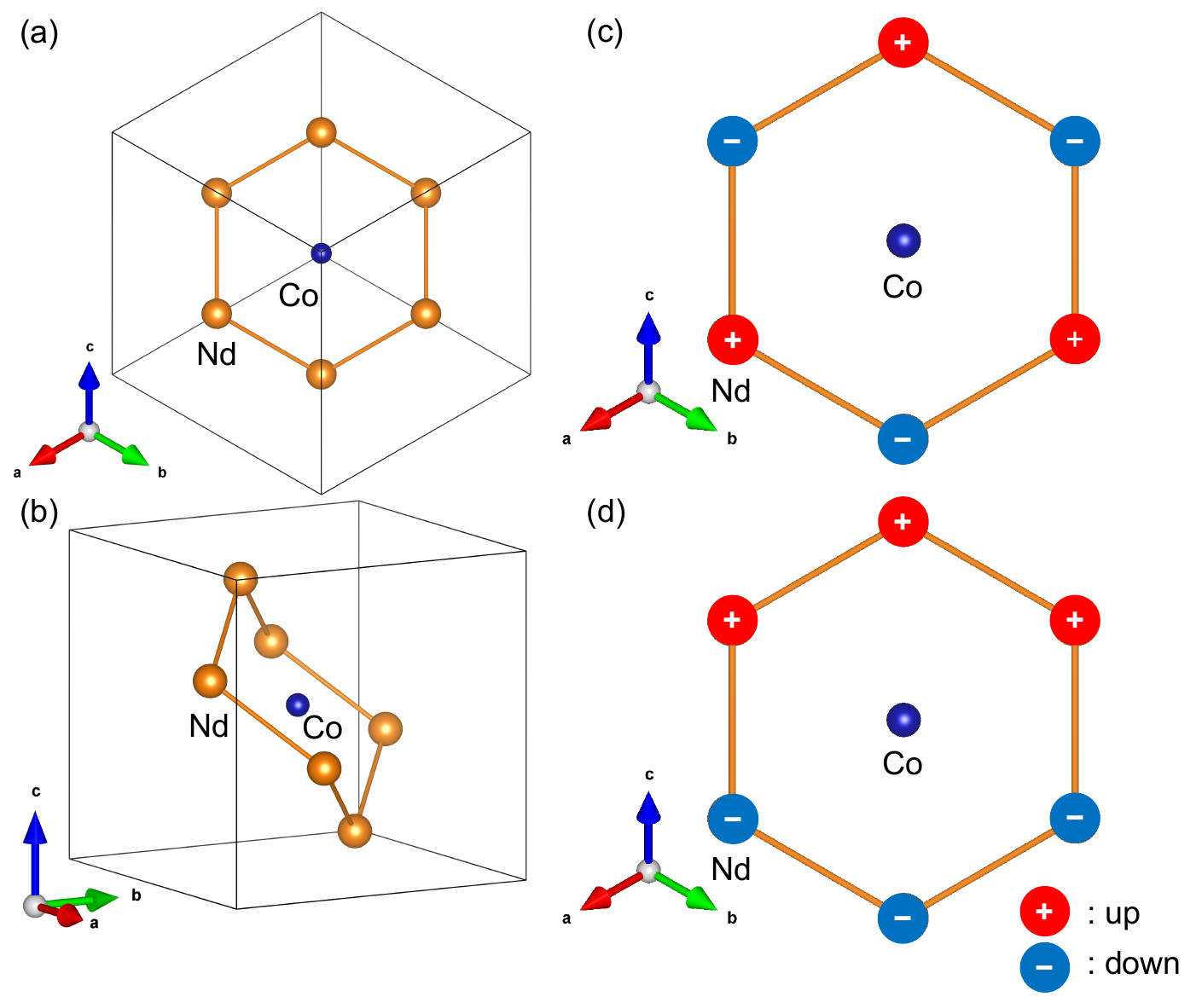}
 \caption{(Color Online) (a,b) Positions of the Co atom and its six nearest-neighbor Nd ions in the cubic unit cell. (c,d) Two candidate alignments of Nd magnetic moments that cancel out the hyperfine field at the Co site. Because the orientation of the moments cannot be determined from NQR measurements, the moments are depicted by $+$ and $-$ symbols. Alignment (c) is compatible with simple antiferromagnetic nearest-neighbor coupling, whereas alignment (d) can arise from more complex magnetic interactions.}
 \label{fig:magstr}
\end{figure}

We now consider the possible magnetic structures that could result in the cancellation of hyperfine fields at the Co sites. The Co sites have local $. 3 m$ symmetry with a threefold rotoinversion axis. Viewed along this axis, the six nearest-neighbor Nd ions form a regular hexagon around each Co atom, as illustrated in Figs.~\ref{fig:magstr}(a,b).

For the hyperfine fields to cancel out at the Co sites, the Nd magnetic moments at symmetrically equivalent positions must align in opposite directions. The possible alignments of the Nd moments are shown in Figs.~\ref{fig:magstr}(c,d). From our NQR results, we can infer only the relative alignment of the nearest-neighbor Nd moments surrounding each Co site. Thus, the moments are represented by $+$ and $-$ symbols.

In the related compounds Nd$T_{2}$Zn$_{20}$ ($T$~=~Co, Rh, and Ir), neutron scattering experiments have determined the propagation vector ${\bf k} = (1 / 2, 1 / 2, 1 / 2)$~\cite{Yamamoto2022,YamamotoPhD}, although the detailed direction of the ordered moments remains unresolved. In NdRh$_{2}$Zn$_{20}$, the representation analysis suggests that the magnetic structure belongs to the $\Gamma_{6}$ representation with a sequence of up-up-down-down ferromagnetic planes along the $[1 1 1]$ direction. However, such an arrangement would result in all nearest-neighbor Nd moments pointing in the same direction around a Co site, which is inconsistent with the absence of NQR spectral changes observed here. Therefore, it is plausible that the magnetic structure changes upon Ga substitution.

According to our representation analysis using the SARAh program~\cite{Wills2000}, several magnetic structures are compatible with the local alignment pattern shown in Fig.~\ref{fig:magstr}(c) for propagation vectors such as ${\bf k} = (0, 0, 0)$, $(1 / 2, 0, 0)$, and $(1 / 2, 1 / 2, 1 / 2)$. As in NdCo$_{2}$Zn$_{20}$, even if the propagation vector remains as ${\bf k} = (1 / 2, 1 / 2, 1 / 2)$, the magnetic unit cell contains eight conventional unit cells, making it difficult to determine the full magnetic structure using NQR alone. Therefore, from a local viewpoint, the plausible magnetic structures can be constrained such that the internal magnetic field at the Ga site is canceled out. Complementary studies using $^{69,71}$Ga-NMR and neutron scattering are necessary to fully elucidate the magnetic structure of this system.

The alignment shown in Fig.~\ref{fig:magstr}(c) can be explained solely by antiferromagnetic interactions between the nearest-neighbor Nd moments. Although magnetic frustration typically arises in diamond structures owing to competing interactions between the first- and second-nearest neighbors, such competition appears to be absent in NdCo$_{2}$Zn$_{18}$Ga$_{2}$.

This interpretation is consistent with bulk measurement results indicating that Ga doping enhances the antiferromagnetic interactions among Nd moments, stabilizing the antiferromagnetic ordering and increasing $T_{\rm N}$ from 0.53~K to 1.5~K. The magnetic entropy at $T_{\rm N}$, which was $0.5 R \ln 2$ in NdCo$_{2}$Zn$_{20}$, increased to $0.8 R \ln 2$ in NdCo$_{2}$Zn$_{18}$Ga$_{2}$, supporting the removal of magnetic frustration by Ga substitution.

By contrast, the alignment shown in Fig.~\ref{fig:magstr}(d) is also consistent with the NQR results. In this case, interactions between the nearest-neighbor Nd moments may involve both ferromagnetic and antiferromagnetic couplings, implying a more complex magnetic structure.

If the two-channel Kondo effect is present, the associated magnetic fluctuations may be detectable. For NdCo$_{2}$Zn$_{20}$, the electrical resistivity $\rho$ follows the two-channel Kondo model $\rho \sim a_{1} / (1 + a_{2} T_{0} / T)$ for $T_{\rm N} < T < 4$~K with $T_{0} = 1.06$~K~\cite{Yamamoto2019,Tsuruta2015}, indicating characteristic magnetic fluctuations. However, this behavior disappeared upon Ga substitution. Above $T_{\rm N}$, the magnetic excitations were dominated by exchange-coupled $4f$ moments. Below $T_{\rm N}$, the magnetic excitations exhibited a temperature dependence similar to that of magnons. At the lowest temperatures, no additional magnetic excitations were observed, and only the contribution of the conduction electrons was observed. Although systems with a $\Gamma_{6}$ doublet ground state may exhibit a residual entropy of $0.5 R \ln 2$~\cite{Hotta2017}, no theoretical model currently accounts for the behavior of other physical quantities. Furthermore, the phase transition becomes second-order with Ga substitution and is well described by mean-field calculations. Combined with the absence of anomalous features in $\rho$ and $1 / T_{1}$ near $T_{\rm N}$, these results suggest that the magnetic fluctuations associated with the two-channel Kondo effect are absent in the present system. Future microscopic studies, such as NMR measurements of samples with diluted Nd concentrations to control intersite interactions, are required to clarify this issue.

In summary, we performed $^{59}$Co-NQR studies on NdCo$_{2}$Zn$_{18}$Ga$_{2}$ to investigate its magnetic properties from a microscopic perspective. The extracted NQR parameters revealed a significant increase in $\nu_{\rm Q}$ with Ga substitution, indicating changes in the EFG at the Co sites. $1 / T_{1}$ exhibited a clear kink at $T_{\rm N} = 1.5$~K, providing microscopic evidence of the antiferromagnetic transition. Notably, the NQR spectra showed no changes across $T_{\rm N}$, indicating that the hyperfine fields cancelled each other out at the Co sites. We proposed two possible Nd moment alignments that could account for the cancellation of the hyperfine fields. Further complementary studies using neutron scattering are required to determine the complete magnetic structure. One of the proposed alignments (Fig.~\ref{fig:magstr}(c)) can be explained solely by antiferromagnetic interactions among nearest-neighbor Nd ions. In this case, it is suggested that the magnetic frustration is removed by Ga substitution, leading to the stabilization of antiferromagnetic order and the enhancement of $T_{\rm N}$. Future investigations should focus on magnetic dynamics through NQR/NMR relaxation rate measurements and the evaluation of magnetic correlations.

{\bf Acknowledgments} This work was financially supported by grants-in-aid from MEXT/JSPS of Japan [Grant No. JP26707017, No. JP15H05886 (J-Physics), No. JP18H01182, No. JP23H04870, No. JP23H04871, No. JP23K03302, and No. JP24K00574], and The Thermal \& Electric Energy Technology Foundation (Grant No. 007 in 2020). We thank the Research Facility Center for Science and Technology of Kobe University for supplying liquid helium and nitrogen.

\end{document}